\documentstyle [12pt, epsf] {article}

\catcode`@=12
\newcommand{\be}{\begin{equation}}
\newcommand{\ee}{\end{equation}}
\newcommand{\bea}{\begin{eqnarray}}
\newcommand{\eea}{\end{eqnarray}}
\newcommand{\nn}{\nonumber}
\def\spur{\not\! \,}
\def\B{B_d^0}
\def\Bb{\bar{B}_d^0}

\def\as {\alpha_s}

\def\d{\delta}

\def\l{\lambda}

\def\p{\pi}
\def\as{\alpha_s}

\def \drho{\bar \rho}
\def \deta{\bar \eta}
\def\etal{ {\em et al.}}

\begin{document}
\vspace{0.5in}
\oddsidemargin -.1 in
\newcount\sectionnumber
\sectionnumber=0

\def\lsim{\mathrel{\vcenter{\hbox{$<$}\nointerlineskip\hbox{$\sim$}}}}
\thispagestyle{empty}

\vskip.5truecm
\vspace*{0.5cm}

\begin{center}
{\Large \bf 
\centerline{Can there be any  new physics in  $b \to d$ penguins }}
\vspace*{0.5cm}
{Anjan K. Giri$^1$,
Rukmani Mohanta$^{1,2}$}
\vskip0.3cm
{\it $^1$Physics Department, 
Technion-Israel Institute of Technology,
Haifa 32000, Israel}
\\
{\it $^2$School of Physics, University of Hyderabad, Hyderabad - 500 046, 
India}\\
\vskip0.5cm
\bigskip
\begin{abstract}
We analyze the possibility of observing new physics effects in the 
$b \to d$ penguin amplitudes. For this purpose, we consider the 
decay mode $\B \to K^0 \bar K^0 $, which has only $b \to d$ penguin
contributions. Using the QCD factorization approach, we find 
very tiny CP violating effects in the standard model for this process. 
Furthermore, we show that the minimal supersymmetric standard 
model with $LR$ mass insertion and R-parity
violating supersymmetric model can provide 
substantial CP violation effects. Observation of
sizable CP violation in this mode would be a clear
signal of new physics.

\end{abstract}
\end{center}

\thispagestyle{empty}
\newpage

\section{Introduction}

Recently there have been a lot of interests to look for new physics effect
beyond the standard model (SM). The recent measurement of the indirect
CP violating parameter $S_{\phi K_S}$ in the decay
mode $\B \to \phi K_S$, which is a pure $b \to s\bar s s $ penguin 
induced process, may provide the first indication 
of new physics beyond the SM \cite{tom1}.
Within the SM,  the mixing induced CP asymmetry in the
$\B \to \phi K_S$ mode is expected to be equal to that of
$\B \to \psi K_S$ \cite{gross1} within a correction of 
${\cal O}(\lambda^2)$. The most recent updated data on
$S_{\phi K^0}$ by BABAR \cite{babar04} agrees within one standard deviation 
with the value of  $S_{\psi K_S}$ whereas, the Belle data 
\cite{belle04} has  about $2 \sigma$ deviation. Therefore, the
presence of new physics (NP) in this mode has not yet been ruled out 
from the available data.
In principle, the new physics can affect either 
the $\B -  \Bb $ mixing or the decay amplitudes. 
Since the new physics effect in the mixing 
can affect equally to both the cases the above deviation may be 
attributed to the decay amplitude of $\B \to \phi K_S$ or more generally
to the $b \to s $ penguin amplitudes. The next obvious question is:
Do the $b \to d$ penguin amplitudes also have significant new physics 
contribution. The present data does not provide any conclusive answer
to it. The obvious example is the $\B \to \pi \pi $ processes, which receive 
contribution from $b \to u$ tree and from $b \to d$ penguin diagrams.
The charge averaged branching ratios of all the three processes 
$\B \to \pi^+ \pi^-$, $\B \to \pi^0 \pi^0 $ and $ B^\pm
\to \pi^\pm \pi^0 $ \cite{pdg04} and the CP violating parameters $C_{\pi \pi}$
and $S_{\pi \pi}$ in $\B \to \pi^+ \pi^-$ process
\cite{pipi} have already been measured. 
The present situation is: the measured branching ratio for the color allowed
process  $\B \to \pi^+ \pi^-$ is about two times
smaller than the QCD factorization calculation and the measured
$\B \to \pi^0 \pi^0 $ color suppressed branching ratio is about
six times larger than the corresponding QCD factorization calculations
\cite{beneke1}.
Thus the discrepancy between the theoretical and the measured quantities
imply the following two consequences.

$\bullet $ The QCD factorization may not be a very successful theory for
the charmless $B$ decays.

$ \bullet $ There may also be significant new physics effect in the $b \to
d$ penguins as speculated in $b \to s$ penguins.

Recently Buras \etal ~\cite{buras04} have shown that the observed 
$B \to \pi \pi $ data
can be explained in the standard model if one includes the  large 
nonfactorizable contributions. 
In this paper we would like to look into the second possibility i.e., 
the existence of any new physics in $b \to d $ penguin amplitudes 
and indeed if it does,
could it be detectable. For this purpose, we consider the decay mode
$\B \to K^0 \bar K^0$ which has only $b \to d $ penguin contribution.
The significance of this decay mode is that it
originates from $ b \to d \bar s s $ penguins with dominant contributions 
coming from the QCD penguins. If one assumes that the penguin topology
is dominated by internal {\it top} quark, the CP asymmetry parameters
would  vanish in the SM. However, as pointed out by Fleischer \cite{fl}, 
contribution from penguins with internal {\it up} and 
{\it charm} quark
exchanges are expected to yield nonzero CP asymmetry in $\B \to K^0 
\bar K^0$ mode. Thus, the study of CP asymmetries in this mode may
provide an interesting testing ground to 
explore new physics effects. The CP averaged branching ratio 
has recently been
measured by  the BABAR collaboration \cite{babar4} 
\be
{\cal B}(\B \to K^0 \bar K^0)=\left (1.19_{-0.35}^{+0.40}\pm 0.13 \right )
\times 10^{-6}\;,
\ee
which agrees with the SM predictions \cite{beneke1}. 
Although, the measured branching ratio
does not provide any indication for a possible new physics effect,
the measurements of CP violation parameters in near future will certainly
establish/rule out the possibility of new physics in the $b \to d $ 
penguin amplitudes. This decay mode has recently been analyzed by
Fleischer and Recksiegel \cite{fle04} in the SM. They have shown that this
channel may be characterized through a surface in the observable space
from which one can extract the relevant information. In this paper we 
consider the impact of new physics effect on the CP violation parameters.
We show that the minimal supersymmetric model with $LR$ mass insertion
and the supersymmetric model with R-parity violation can provide
significant CP violation effect, the observation of which would be a
clear signal of new physics.
  
The paper is organized as follows. In section 2, we discuss the basic 
formalism of CP violation. Using QCD factorization approach,
we estimate the CP averaged branching ratio and CP violating parameters 
in the SM. The effects of new physics on the CP violating parameters are
discussed in section 3. The contributions arising from minimal supersymmetric 
standard  model with mass insertion approximation  
and from R-parity violating supersymmetric 
model are discussed in subsections 3.1 and 3.2 respectively.
Our conclusion is presented in section 4.

\section{CP violating parameters in the Standard Model}

We first present a very general treatment of the CP violating parameters.
The time dependent CP asymmetry for $\B \to K^0 \bar K^0$
can be described by 
\be
a_{K K}(t)=C_{K K}\cos \Delta M_{B_d} t+ S_{K K}
\sin \Delta M_{B_d} t\;,
\ee
where 
\be
C_{K K}=\frac{1-|\lambda|^2}{1+|\lambda|^2},~~~~
S_{K K}=-\frac{2 {\rm Im(\lambda)}}{1+|\lambda|^2}.\label{cv}
\ee
In the above expression $\lambda$ corresponds to
\be
\lambda=\frac{q}{p}\frac{A(\Bb \to  K^0 \bar K^0)}{A(\B \to 
K^0 \bar K^0)}\;,
\ee
where, $q$ and $p$ are the mixing parameters defined as
\be
\frac{q}{p}= \sqrt {\frac{M_{12}^* -\frac{i}{2} \Gamma_{12}^*}
 {M_{12} -\frac{i}{2} \Gamma_{12}}}\;.\label{mix}
\ee
The off diagonal element of the mass matrix is given by the matrix 
element of the  $\Delta B=2$ transition as
\be
\langle B_d^0 | {\cal H}_{\Delta B =2}|\Bb \rangle = M_{12}-\frac{i}{2}
\Gamma_{12}\;.
\ee
In the standard model, the box diagrams are dominated by the $W$-boson 
and {\it top} 
quark in the loop, as a result of which, one obtains (ignoring
terms of ${\cal O}(\Gamma_{12}/M_{12}))$
\be
\frac{q}{p}=\frac{V_{tb}^* V_{td}}{V_{tb} V_{td}^*}\simeq e^{-2i \beta}\;.
\ee
The amplitude for the decay mode $\B \to K^0 \bar K^0 $, which
receives  dominant contribution in the SM from QCD penguins  
with {\it top} quark in the loop can
be written as
\be
A( \Bb \to K^0 \bar K^0) = V_{tb}V_{td}^*~ P_t \;,
\ee  
where $V_{ij}$ are the CKM matrix elements which provide the weak 
phase information and 
$P_t$ is the penguin amplitude arising from the matrix elements of
the four quark operators of the effective Hamiltonian. The amplitude for
the corresponding CP conjugate process is given as
\be
A( \B \to K^0 \bar K^0) = V_{tb}^*V_{td}~ P_t\;. 
\ee  
Thus one gets
 \be
\lambda =\left (\frac{V_{tb}^* V_{td}}{V_{tb} V_{td}^*}\right )
\left (\frac{V_{tb} V_{td}^*}{V_{tb}^* V_{td}}\right )=1\;,
\ee
and hence
\be
C_{KK}=S_{KK}=0\;.
\ee
Thus if the measured CP violating asymmetries in $B^0 \to K^0 \bar K^0$
deviates significantly from zero it would be a clear signal of new physics.
However, the decay amplitude
also receives some contribution from the internal {\it up} and 
{\it charm} quarks in the loop. Therefore, the CP violating
parameters 
may not be zero identically. Now including the effects of $u, c,t$
quarks in the loop and using CKM unitarity one can write the decay
amplitude as
\bea
A(\Bb \to K^0 \bar K^0) = \lambda_u P_{ut} + \lambda_c P_{ct}
=\lambda_u P_{ut}\left [1-re^{i(\delta +\gamma)}\right ]\;,\label{amp}
\eea  
where $\lambda_q= V_{qb}V_{qd}^*$, $P_{qt}=P_q-P_t$ are the penguin 
amplitudes,  $\delta= \delta_{ct}-\delta_{ut}$ is the relative strong 
phase between them and  $\gamma$ is the weak phase. The parameter $r$
is defined as 
\be
r=\frac{1}{R_b} \left | \frac{P_{ct}}{P_{ut}} \right |\;,~~~~~
{\rm with}~~~~ R_b=\left (1-\frac{\l^2}{2}\right ) 
\frac{1}{\l}\left |\frac{V_{ub}}{V_{cb}}
\right |=\sqrt{\drho^2 +\deta^2}\;.
\ee 
Thus one obtains the CP asymmetries as
\bea
S_{KK} &=&- \frac{\sin 2 \alpha + 2r \cos \delta \sin(2 \beta+\gamma)
-r^2 \sin 2\beta }{1+r^2-2 r \cos \delta \cos \gamma}\;,\nn\\
C_{KK}&=& \frac{-2r \sin \delta \sin \gamma}{1+r^2-2r \cos \delta
\cos \gamma}\;,\label{cpa}
\eea
where $\alpha$, $\beta$ and $\gamma$ are the three angles of the
unitarity triangle. Thus, to know the precise value of the CP
violating asymmetries we must know the values of $r$ and $\delta$.
The CP averaged branching ratio for the process is given as
\be
 {\cal B} (\B \to K^0 \bar K^0) =\frac{1}{2}\left [
{\rm Br}(\B \to K^0 \bar K^0)+{\rm Br}(\Bb \to K^0 \bar K^0) 
\right ]\;,
\ee
where the individual branching ratios are given as
\be
{\rm Br}(\B \to K^0 \bar K^0)=\frac{\tau_{B^0} |p_{c.m}|}{8 \p m_B^2}~
\left |A(\B \to K^0 \bar K^0)\right |^2\;.
\ee 
We now use the QCD factorization approach  to calculate the 
branching ratio and CP asymmetry parameters.
The effective Hamiltonian describing the process under consideration is
\be 
{\cal H}_{eff}=
\frac{G_F}{\sqrt{2}}
V_{qb}V_{qd}^* \biggr [\sum_{j=3}^{10}C_j O_j+C_g O_g
 \biggr],
\ee
where $q=u,~c$.  $O_3, \cdots, O_{6}$ and $O_7,
\cdots, O_{10}$ are the standard model QCD and electroweak penguin operators
respectively, and $O_{g}$ is the gluonic magnetic penguin operator.
The values of the Wilson coefficients at the scale $\mu \approx m_b$
in the NDR scheme are given
in Ref. \cite{buca96} as
\bea
&&C_3=0.014\;,~~~~~C_4=-0.035\:,~~~~~C_5=0.009\;,~~~~C_6=-0.041\;,
~~~~C_7=-0.002\alpha\nn\\
&&C_8=0.054 \alpha\;,~~~~~
C_9=-1.292\alpha\;,~~~~C_{10}=0.263 \alpha\;,~~~~C_{g}=-0.143\;.
\eea
In the QCD factorization approach \cite{beneke1},  
the hadronic matrix elements can be represented in the form
\be
\langle K^0 \bar  K^0 |O_i |\B \rangle =
\langle K^0  \bar K^0 |O_i |\B \rangle_{\rm fact}\biggr[
1+\sum r_n \alpha_s^n + {\cal O}(\Lambda_{\rm QCD}/m_b)
\biggr]\;,
\ee
where $\langle K^0 \bar K^0 |O_i |\B \rangle_{\rm fact}$ denotes the 
naive factorization result and $\Lambda_{\rm QCD} \sim 225$ MeV is 
the strong interaction scale. The second and third terms in the square
bracket represent higher order $\alpha_s$ and $\Lambda_{\rm QCD}/m_b$
corrections to the hadronic matrix elements. 

In the heavy quark limit the decay amplitude for the $\B \to K^0 \bar K^0 $
process, arising from the penguin diagrams is given  as
\bea
A(\Bb \to K^0 \bar K^0) = \frac{G_F}{\sqrt 2}\sum_{q=u,c} 
V_{qb}V_{qd}^* \biggr[a_4^q-\frac{a_{10}^q}{2}+
a_{10a}^q+r_{\chi}\left (a_6^q-\frac{a_8^q}{2}
+a_{8a}^q \right )\biggr]X\;,\label{smamp}
\eea
where $X$ is the factorized matrix element. Using the form factors and decay
constants defined as \cite{bsw}
\bea
\langle K^0(p_K) |\bar s \gamma ^\mu b |\Bb (p_B) \rangle &=&
\biggr[(p_B+p_K)^\mu-\frac{m_B^2-m_K^2}{q^2} q^\mu \biggr] F_1(q^2)\nn\\
&+&
\frac{m_B^2-m_K^2}{q^2}q^\mu F_0(q^2)\;,\nn\\
\langle \bar K^0 (q ) |\bar d \gamma ^\mu  \gamma_5 s | 0 \rangle &=&
-i f_K~ q^\mu\;,
\eea
we obtain 
\bea
 X &=& \langle  K^0 (p_K)| \bar s
\gamma_\mu(1-\gamma_5)b | \Bb (p_B) \rangle
\langle \bar K^0 (q )|\bar d
\gamma^\mu(1-\gamma_5)s|0 \rangle \nn\\
& = & -i f_K F_0(m_K^2)~(m_B^2-m_K^2)
\;.
\eea
The coefficients $a_i^q$'s which contain next to leading order
(NLO) and hard scattering corrections are given as \cite{yang1,du1}
\bea
a_4^q &= & C_4+\frac{C_3}{N}+\frac{\alpha_s}{4 \pi} \frac{C_F}{N}
\biggr[ C_3 \left [F_K + G_K(s_d) + G_K(s_b) \right ]
\nn\\
&+&C_1 G_K(s_q) +(C_4+C_6) \sum_{f=u}^b G_K(s_f)
+C_{g} G_{K, g}\biggr]\;,\nn\\
a_6^q &= & C_6+\frac{C_5}{N}+\frac{\alpha_s}{4 \pi} \frac{C_F}{N}
\biggr[ C_3 \left [ G_K^\prime(s_d) + G_K^\prime(s_b) \right ]
+C_1 G_K^\prime(s_q) \nn\\
&+&(C_4+C_6) \sum_{f=u}^b G_{K}^\prime(s_f)
+C_{g} G_{K, g}^\prime \biggr]\;,\nn\\
a_8^q &=&  C_8+\frac{C_7}{3}\;,\nn\\
a_{8a}^q &=&\frac{\alpha_s}{4 \pi} \frac{C_F}{N}
\biggr[ (C_8+C_{10})\frac{3}{2}\sum_{f=u}^b e_f G_K^\prime(s_f)
+C_9 \frac{3}{2} \left [ e_dG_K^\prime(s_d) +e_b  G_{K}^\prime
(s_b) \right ]\biggr]\;,\nn\\
a_{10}^q &=&  C_{10}+\frac{C_9}{N}+\frac{\alpha_s}{4 \pi}
\frac{C_F}{N}C_9 F_K \;, \nn\\
a_{10a}^q &= & \frac{\alpha_s}{4 \pi} \frac{C_F}{N}
\biggr[( C_8+C_{10}) \frac{3}{2}
\sum_{f=u}^{b}e_f G_K(s_f)
 +  C_9 \frac{3}{2}\left [e_d G_K(s_d) +e_b G_K(s_b) \right ]
\biggr]\;,\label{qcd}
\eea
where $q$ takes the values $u$ and $c$, $N=3$, is the number of
colors, $C_F=(N^2-1)/2N$. The internal quark mass in the
penguin diagrams enters as  $s_f=m_f^2/m_b^2$. 
The other parameters in (\ref{qcd}) are given as
\bea
F_K &=& -12 \ln \frac{\mu}{m_b}-18+f_K^I +f_K^{II}\;,\nn\\
f_K^I &= & \int_0^1 dx~ g(x) \phi_K(x)\;,~~~~
g(x) =  3 \frac{1-2x}{1-x}\ln x -3i \pi\;, \nn\\
f_K^{II} &= & \frac{4 \pi^2}{N} \frac{f_K f_B}{F_0^{B \to K}(0) m_B^2}
\int_0^1 \frac{dz}{z} \phi_B(z) \int_0^1 \frac{dx}{x} \phi_K(x) 
\int_0^1 \frac{dy}{y} \phi_K(y)\;, \nn\\
G_{K,g} &=& - \int_0^1 d x \frac{2}{1-x} \phi_K(x)\;,\nn\\
G_K(s) &=& \frac{2}{3}-\frac{4}{3}{\rm ln}\frac{\mu}{m_b}
+4\int_0^1 dx~ \phi_K(x) \nn\\
&\times & \int_0^1 du~u (1-u) \ln\left [ s-u(1-u)(1-x)-i \epsilon \right ]\;,
\nn\\
G_{K,g}^\prime &=& -\int_0^1 dx \frac{3}{2} \phi_K^0(x)=-\frac{3}{2}\;,\nn\\
G_K^\prime(s) &=& \frac{1}{3}-{\rm ln}\frac{\mu}{m_b}
+3\int_0^1 dx~ \phi_K^0(x) \nn\\
&\times & \int_0^1 du~u (1-u) \ln\left [ s-u(1-u)(1-x)-i \epsilon \right ]\;.
\eea
The light cone distribution amplitudes (LCDA's) at twist two order 
are given as
\bea
&&\phi_B(x) = N_B x^2(1-x)^2 {\rm exp}\biggr(-\frac{m_B^2 x^2}{2
\omega_B^2}\biggr) \;,\nn\\
&&\phi_{K}(x)=  6x(1-x)\;,~~~~\phi_K^0(x)=1\;,
\eea
where $N_B$ is the normalization factor satisfying
$ \int_0^1 dx \phi_B(x)=1$ and $\omega_B=0.4$ GeV. The quark masses appear in
$G(s)$ are pole masses and we have used the following values (in GeV)
in our analysis
\begin{eqnarray*}
m_u=m_d=m_s=0, ~~~~m_c=1.4~~~~m_b=4.8.
\end{eqnarray*}
$r_\chi=2 m_K^2/(m_b-m_s)(m_s+m_d)$ denotes the chiral enhancement factor.
It should be noted that the quark masses in the chiral enhancement factor
are running quark masses and we have used their values  at the $b$ 
quark mass scale
 as $m_b(m_b)$=4.4 GeV, $m_s(m_b)$=90 MeV and $m_d(m_b)=6.4$ MeV.

For numerical evaluation we have used the following input parameters. 
The value of the form factor 
at zero recoil is taken as $F_0(0)=$ 0.38, 
and its value at $q^2=m_K^2$ can be obtained 
using simple pole dominance ansatz \cite{bsw} as 
$F_0(m_K^2)=$ 0.383. 
The  values of the decay constants are as $f_{K}=$ 0.16 GeV 
and $f_B=0.19 $ GeV,
the particle masses and the lifetime of 
$\B$ meson
 $\tau_{B^0}=1.536$ ps are taken from \cite{pdg04}. 
Thus we obtain the amplitude (in units of $10^{-2}$)
\bea
A(\Bb \to K^0 \bar K^0)&=& -\frac{G_F}{\sqrt 2}~\biggr[
V_{ub}V_{ud}^*\left (13.56 + i~4.59 \right )+V_{cb}V_{cd}^*
\left (14.98+i~
2.06 \right )\biggr]\nn\\
&=& -\frac{G_F}{\sqrt 2}~\biggr[
V_{ub}V_{ud}^* \left (14.32~ e^{i 18.7^\circ}
\right )+V_{cb}V_{cd}^*\left (15.12~ e^{i 7.83^\circ}\right )
\biggr]\;.\label{eq:ampsm}
\label{kl1}
\eea
We use the values of the CKM matrix elements at the $1 \sigma$ CL 
in the Wolfenstein
parameterization from Ref. \cite{cha04} as 
\begin{eqnarray}
\l=0.2265_{-0.0023}^{+0.0025}\;,~~~A=0.801_{-0.020}^{+0.029}\;,~~~~
\drho=0.189_{-0.070}^{+0.088}\;,~~~\deta=0.358_{-0.042}^{+0.046}\;,
\label{ckm1}
\end{eqnarray}
which correspond to  the angles of the CKM unitarity triangle
\begin{eqnarray}
\sin 2 \alpha = -0.14_{-0.41}^{+0.37}\;,
~~~\beta= \left (23.8_{-2.0}^{+2.1}
\right )^\circ\;,~~~~\gamma=\left(62_{-12}^{+10} 
\right )^\circ\;.\label{ckm}
\end{eqnarray}
With these input parameters we obtain the CP averaged branching ratio as
\be
 {\cal B}(\B \to K^0 \bar K^0) = (9.15 \pm 
0.30)\times 10^{-7}\;, \label{br}
\ee
which is slightly below the central experimental value (1). 
Since, we
have used the LCDA's at twist two order our predicted result is slightly
lower than that of Ref. \cite{beneke1} where they have included the
twist three power corrections in the distribution amplitudes.
From Eqs. (\ref{amp}) and (\ref{eq:ampsm}) one can obtain
\be
r \approx 2.6 ~~~{\rm and}~~~\d \approx 11^\circ\;.\label{rp}
\ee 
With these values we get the CP asymmetry parameters in the
SM as
\be
S_{KK}=0.061 ~~~~~~C_{KK}=-0.163\;.
\ee
By allowing the CKM matrix elements to vary within
their $1 \sigma$ range as given in Eqs. (\ref{ckm1}) and
(\ref{ckm}), we obtain the correlation  between 
$S_{KK}$ and $C_{KK}$ in the SM as shown in Figure-1, which gives
the constraints 
\be
0.02 \leq S_{KK} \leq 0.13\;,~~~~~~-0.17 \leq C_{KK} \leq -0.15\;.
\ee
Thus, if the measured values of CP asymmetry parameters 
will be outside the above ranges,  would be a 
clear sign of new physics. 
\begin{figure}[htb] 

 \centerline{\epsfysize 3.0 truein \epsfbox{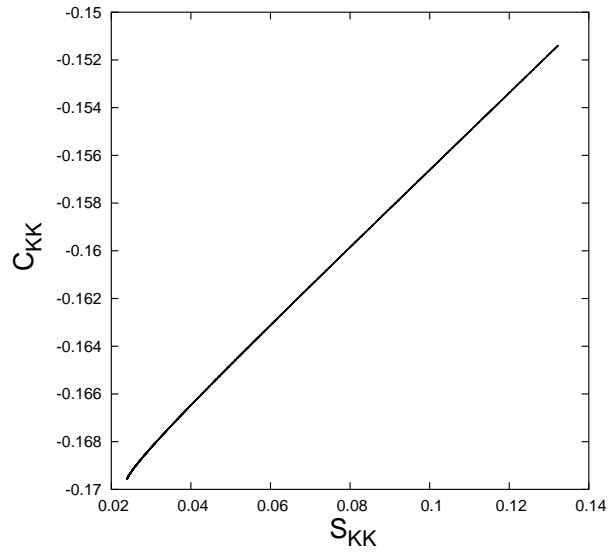}}
\caption{The  correlation plot between $S_{KK}$ and $C_{KK}$
for the $\B \to K^0 \bar K^0$ process in the SM, where we have used
$r=2.6$, $\delta=11^\circ$ and the CKM parameters are
varied within the range as given
in Eqs. (\ref{ckm1}) and (\ref{ckm}).}
\label{cor}
 \end{figure}


\section{New Physics effects on the CP violating parameters}
 
Here, we consider the effect of new physics on the CP
violating parameters. Because of the new
physics contributions, the CP asymmetry parameters (\ref{cpa})
become modified. In principle, the new physics can affect
either the $B_d^0-\Bb$ mixing or the decay amplitudes. Let us first
investigate its effect in the mixing phenomena. In the presence of 
new physics, there are additional contributions to the mixing parameters
arising from the new box diagrams. These contributions to the $\Delta B=2$
transitions are often parametrized as \cite{yg}
\be
\sqrt{\frac{M_{12}}{M_{12}^{SM}}}=r_m e^{i \theta_m}\;,
\ee
where $M_{12}=M_{12}^{SM}+M_{12}^{NP} $ is the off diagonal
element of the mass matrix, contains contribution both from
the SM and from new physics. Hence, the ratio of the mixing 
parameters $q/p$ as given in Eq.(\ref{mix}) becomes

\be
\frac{q}{p}\sim e^{-2i (\beta+\theta_m)}\;.
\ee
Thus, in the presence of new physics, the mixing induced CP asymmetry
in $B_d^0 \to \psi K_S$ can be given as
\be
S_{\psi K_S}=\sin(2\beta+2\theta_m)
\ee
However, since the present world average on
the measurement of $S_{\psi K_S}=0.726 \pm 0.037$ \cite{ligeti}
agrees quite well with the SM prediction 
$S_{\psi K_S}=0.715_{-0.045}^{+0.055}$
\cite{smb}, we do not consider the effect of NP in
mixing in our analysis. 

Now we consider the effect of new physics on the CP violating parameters
arising from the new contribution  
to the standard model decay amplitude (\ref{amp}).
In the presence of new physics the amplitude can be written as
\bea
A(\Bb \to K^0 \bar K^0)=A_{SM}+A_{NP}
= \lambda_u P_{ut}\left [1-re^{i(\delta+\gamma)}+ 
r_{NP} e^{i \theta_N} e^{i \delta_N} \right ]
\;,\label{npamp}
\eea
where $r_{NP}=|A_{NP}/\lambda_u P_{ut}|$, $\theta_N$ and $ {\delta_N}$
are the relative weak and strong phase between them. 
The amplitude for the corresponding CP conjugate process can be
obtained by changing the sign of the weak phases.

Thus the CP asymmetry parameters (\ref{cv}) become 
\be
S_{KK}^{NP}= -\frac{X}{Y}\;,~~~~~~{\rm and}~~~~~~~
C_{KK}^{NP}=-\frac{Z}{Y}\;,
\ee
where
\bea
X &=&  \sin 2 \alpha +2r \cos \delta \sin (2 \beta+\gamma)
- r^2 \sin 2 \beta +2 r_{NP} \cos \delta_N \sin (2 \alpha+\theta_N)\nn\\
&&-
2 r r_{NP} \cos (\delta -\delta_N)\sin (\theta_N-(2 \beta+\gamma))
+r_{NP}^2 \sin (2 \alpha+2 \theta_N)\;,\nn\\
Y&=& 1+r^2+r_{NP}^2 -2r \cos \delta \cos \gamma+ 2 r_{NP} \cos \delta_N 
\cos \theta_N \nn\\
&&-2 r r_{NP} \cos(\delta -\delta_N) \cos (\gamma -\theta_N)
\;,\nn\\
Z &= &2 \biggr[r \sin \delta \sin \gamma - r_{NP} \sin \delta_N \sin \theta_N
+r r_{NP} \sin(\delta - \delta_N) \sin(\gamma- \theta_N )\biggr]\;.
\eea

The  branching ratio in the presence of new physics is given as
\be
 {\rm Br}(\Bb \to K^0 \bar K^0)={\rm Br^{SM}}\left (1+\left |\frac{A_{NP}}
{A_{SM}}\right |^2+2 r_{NP}\left |\frac{A_{NP}}
{A_{SM}}\right | \cos
\phi_{N}  \right )\;,\label{br11}
\ee
where ${\rm Br^{SM}}$ denotes the SM branching ratio and
$\phi_N$ is the relative phase between the new physics and standard model
amplitudes.
 
Now, we consider two beyond the standard model scenarios: the minimal
supersymmetric standard model with mass insertion approximation and
R-parity violating supersymmetric model and study their effects
on CP violation parameters in the
following subsections.

\subsection{Contribution from minimal supersymmetric standard 
model with mass insertion approximation}

Here, we analyze the decay process $\B \to K^0 \bar K^0 $, in the minimal
supersymmetric standard model (MSSM) with mass insertion approximation.
This decay mode receives supersymmetric (SUSY) 
contributions mainly from penguin and box diagrams containing
gluino-squark, chargino-squark and 
charged Higgs-top loops. Here, we consider only the
gluino contributions, because
the chargino and charged Higgs loops are expected to be suppressed
by the small electroweak gauge couplings. 
However, the gluino mediated FCNC contributions are 
of the order of strong interaction strength, which may exceed the
existing limits.  Therefore, it is customary to rotate the 
effects, so that the FCNC effects occur
in the squark propagators rather than in couplings and to parameterize them
in terms of dimensionless parameters. Here we work in the usual mass insertion
approximation \cite{hall86, gabb96} where the flavor mixing
$i \to j$ in the down-type squarks associated with $\tilde q_B$ and
$\tilde q_A$ are parametrized by  $(\delta^d_{AB})_{ij}$, with
$A,~B=L,~R$ and $i,j$ as the generation indices. More explicitly
$(\delta^d_{LL})_{ij}
=({V_L^d}^\dagger M_{\tilde d}^2 V_L^d)_{ij}/ m_{\tilde q}^2$, where 
$M_{\tilde d}^2$
is the squared down squark mass matrix and $m_{\tilde q}$ is
the average squark mass.
$V_d$ is the matrix which diagonalizes the down-type quark mass matrix.

The new effective $\Delta B=1$ Hamiltonian relevant for the $B \to K^0 \bar 
K^0$
process arising from new penguin/box diagrams with gluino-squark in the 
loops is given as
\be
{\cal H}_{eff}^{SUSY} = -\frac{G_F}{\sqrt 2} V_{tb}V_{td}^*
\left [\sum_{i=3}^6 \left ( C_i^{NP}O_i+ \tilde C_i^{NP} \tilde O_i
\right )
+C_g^{NP} O_g + \tilde C_g^{NP} \tilde O_g \right ]\;,
\ee
where $O_i$ ($O_g$) are the QCD (magnetic) penguin operators 
and $C_i^{NP}$, $C_g^{NP}$ are the 
new Wilson coefficients. The operators $\tilde O_i$ 
are obtained from $O_i$  by exchanging $L \leftrightarrow
R$.

To evaluate the
amplitude in the MSSM, we have to first determine the Wilson coefficients 
at the $b$ quark mass scale.
At the leading order in mass insertion approximation the new Wilson 
coefficients corresponding to each of the operator at the scale
$\mu \sim \tilde m \sim M_W$ are given as \cite{gabb96,ko04}
\bea 
C_3^{NP} & \simeq & -\frac{\sqrt 2 \alpha_s^2}{4 G_F V_{tb}
V_{td}^* m_{\tilde q}^2}\left ( \delta_{LL}^d \right )_{13}
\left [ - \frac{1}{9}B_1(x) -\frac{5}{9} B_2(x)-\frac{1}{18}P_1(x)
-\frac{1}{2}P_2(x) \right ]\;,\nn\\
C_4^{NP} & \simeq & -\frac{\sqrt 2 \alpha_s^2}{4 G_F V_{tb}
V_{td}^* m_{\tilde q}^2}\left ( \delta_{LL}^d \right )_{13}
\left [ - \frac{7}{3}B_1(x) +\frac{1}{3} B_2(x)+\frac{1}{6}P_1(x)
+\frac{3}{2}P_2(x) \right ]\;,\nn\\
C_5^{NP} & \simeq & -\frac{\sqrt 2 \alpha_s^2}{4 G_F V_{tb}
V_{td}^* m_{\tilde q}^2}\left ( \delta_{LL}^d \right )_{13}
\left [  \frac{10}{9}B_1(x) +\frac{1}{18} B_2(x)-\frac{1}{18}P_1(x)
-\frac{1}{2}P_2(x) \right ]\;,\nn\\
C_6^{NP} & \simeq & -\frac{\sqrt 2 \alpha_s^2}{4 G_F V_{tb}
V_{td}^* m_{\tilde q}^2}\left ( \delta_{LL}^d \right )_{13}
\left [ - \frac{2}{3}B_1(x) +\frac{7}{6} B_2(x)+\frac{1}{6}P_1(x)
+\frac{3}{2}P_2(x) \right ]\nn\\
C_{g}^{NP} &\simeq & 
-\frac{2\sqrt 2 \pi \alpha_s}{2 G_F V_{tb}
V_{td}^* m_{\tilde q}^2}\biggr[ \left ( \delta_{LL}^d \right )_{13}
\left (  \frac{3}{2}M_3(x) -\frac{1}{6} M_4(x)\right )\nn\\
&&\hspace*{1.0 true in}
+ \left ( \delta_{LR}^d \right )_{13}
\left (\frac{m_{\tilde g}}{m_b}\right )
\frac{1}{6}\left (4B_1(x)-\frac{9}{x}B_2(x) \right )\biggr]\;,
\eea
where $x=m_{\tilde g}^2/m_{\tilde q}^2 $. The loop functions appear in these
expressions can be found in Ref. \cite{gabb96}. 
The corresponding $ \tilde C_i$ are obtained from $C_i^{NP}$ by
interchanging $L \leftrightarrow R$. 
It should be noted that the $(\delta_{LR}^d)_{13}$ contribution
is enhanced by $(m_{\tilde g}/m_b)$ compared to the contribution from
the SM and the $LL$ insertion due to the chirality flip from the
internal gluino propagator in the loop.

The Wilson coefficients at low  energy
$C_i^{NP}(\mu \sim m_b)$,  can be obtained from $C_i^{NP}(M_W)$
by using the Renormalization Group (RG) equation as discussed in
Ref. \cite{buca96}, as
\be
{\bf C}(\mu) ={\bf U}_5(\mu, M_W) {\bf C}(M_W)\;,
\ee
where ${\bf C}$ is the $6 \times 1$ column vector of the
Wilson coefficients and
${\bf U}_5(\mu, M_W)$ is the five-flavor $6 \times 6$ evolution matrix.
In the next-to-leading order (NLO), ${\bf U}_5(\mu, M_W)$ is given by
\be
{\bf U}_5(\mu, M_W)=\left (1+\frac{\alpha_s(\mu)}{4 \pi} {\bf J} \right )
{\bf U}_5^{(0)}(\mu, M_W)\left (1-\frac{\alpha_s(M_W)}{4 \pi} {\bf J} 
\right )\;,
\ee
where ${\bf U}_5^{(0)}(\mu, M_W)$ is the leading order (LO)
evolution matrix and ${\bf J}$ denotes the NLO corrections to the evolution. 
The explicit forms of ${\bf U}_5(\mu, M_W)$ and 
${\bf J}$ are given in Ref. \cite{buca96}. 

Since the $O_g$ contribution to the matrix element is $\alpha_s$ order 
suppressed, we consider only leading order RG  effects for
the coefficient $C_g^{NP}$, which is given as \cite{he01}
\be
C_g^{NP}(m_b)= \left ( \frac{\as(m_{\tilde q}}{\as(m_t)} \right )^{2/21}
\left ( \frac{\as(m_t)}{\as (m_b)} \right )^{2/23}\;.
\ee
For the numerical analysis, we fix the SUSY parameter as $m_{\tilde q}
=m_{\tilde g}=500 $ GeV, $\alpha_s(M_W)=0.119$,
$\alpha_s(m_b=4.4$ GeV)=0.221, $\alpha_s(m_t=175$ GeV)=0.107. 
Now substituting the values of the RG evoluted 
Wilson coefficients $C_i^{NP}(m_b)$'s  in Eq. (\ref{qcd}) we obtain
the corresponding $a_i$'s and hence with Eq. (\ref{smamp})
the amplitude. Assuming that all the mass insertion parameters
$ (\delta^d_{AB})_{13}$ have a common weak phase, we obtain the 
fraction of new physics amplitude as defined in Eq. (\ref{npamp})
\bea
r_{NP}\simeq 
0.33 \left (| (\delta^d_{LL} )_{13}| -
| (\delta^d_{RR} )_{13}|\right )+
465.86 \left (| (\delta^d_{LR} )_{13}|
-| (\delta^d_{RL})_{13}|\right )\;.\label{LL}
\eea  
It should be noted that because of the opposite
chiral structure of the currents $O_i$ and $\tilde O_i$,
the $LL$ and $RR$ and also the $LR$ and $RL$
contributions occur with opposite sign.
As seen from Eq. (\ref{LL}), the $LR (RL)$ insertions have 
dominant effect because of the $m_{\tilde g}/m_b$ enhancement. 
We use the limits on the $ (\delta^d_{LL} )_{13}$ and 
$(\delta^d_{LR} )_{13}$ mixing parameters from \cite{ko02}
for $x=1$ as
\be
|(\delta^d_{LL} )_{13}| \leq 0.2~~~~~~~~~
|(\delta^d_{LR} )_{13}|\leq 0.01
\ee
and assume that only one of these gives a dominant SUSY contribution.
This gives $(r_{NP})_{LL} \leq 6.6 \times 10^{-2}$ and
 $(r_{NP})_{LR} \leq 4.66$.
Since the new physics effect due to $LL$ insertion is almost
negligible it will not provide any significant effect on the CP
violating observables. The correlation plot between $C_{KK}^{NP}$ and
$S_{KK}^{NP}$ for $LL$ insertion is shown in Figure-2, where we use
$r=2.6$, $\delta=11^\circ$ as obtained from QCD factorization
analysis, the central values of the CKM weak phases from
(\ref{ckm}), the relative weak phase 
$\theta_N=\pi$ and vary the relative
strong phase $\delta_{N}$ between 0 and $2\pi$. In this case, because of the
negligible new physics contribution one gets  only
tiny CP violating effects. In Figure-3, we present the
correlation plot for $LR$ insertion, where we  have used
$(r_{NP})_{LR} = 4.66$, $0 \leq \delta_{N} \leq 2 \pi$ and
a  representative set of weak phases  $\theta_{N}=\pi,~\pi/2,~
\pi/3,~\pi/4$. For $r$ and $\delta$, we have used the values as
obtained from QCD factorization (\ref{rp}).
As expected, in this case large CP violation can be generated.

The  branching ratio (\ref{br11}) versus $\phi_{N}$ is plotted in Figure-4
for $|A_{NP}/A_{SM}|=2.19$.
One can  see from Figure-4 that the observed data can be easily 
accommodated in minimal supersymmetric standard model with $LR$ 
mass insertion.   

Thus in future, if sizable CP violation effects will be observed in
$\B \to K^0 \bar K^0 $ mode, the minimal supersymmetric standard
model with $LR$ mass insertion may be a strong contender of
new physics to explain the data. 
\begin{figure}[htb] 
 \centerline{\epsfysize 3.5 truein \epsfbox{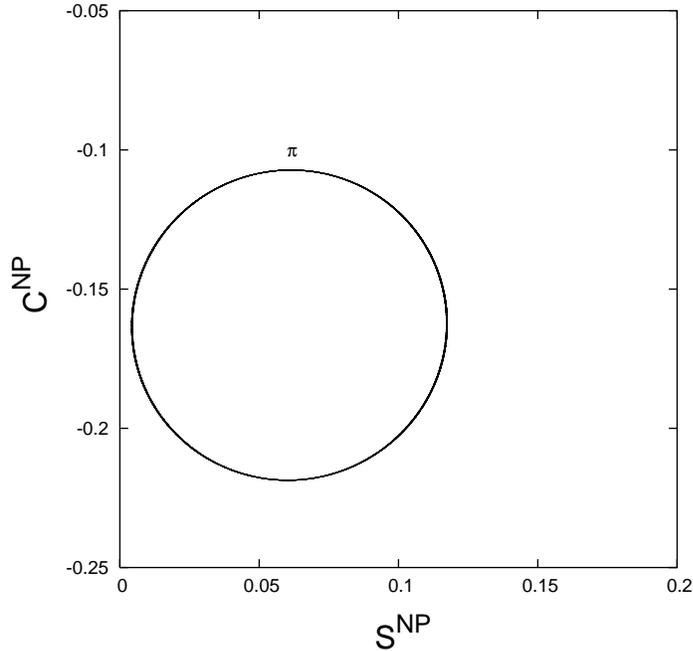}}
\caption{The  correlation plot between $S_{KK}^{NP}$ and $C_{KK}^{NP}$
for the $\B \to K^0 \bar K^0$ process in the MSSM with only $LL$ insertion, 
where we have used $(r_{NP})_{LL}=6.6 \times 10^{-2}$, the weak phase
$\theta_{N}=\pi$, $r=2.6$, $\delta=11^\circ$
and  varied the strong phase
$\delta_{N}$ between 0 and $2 \pi$. }
 \end{figure}

\begin{figure}[htb] 
\centerline{\epsfysize 3.5 truein \epsfbox{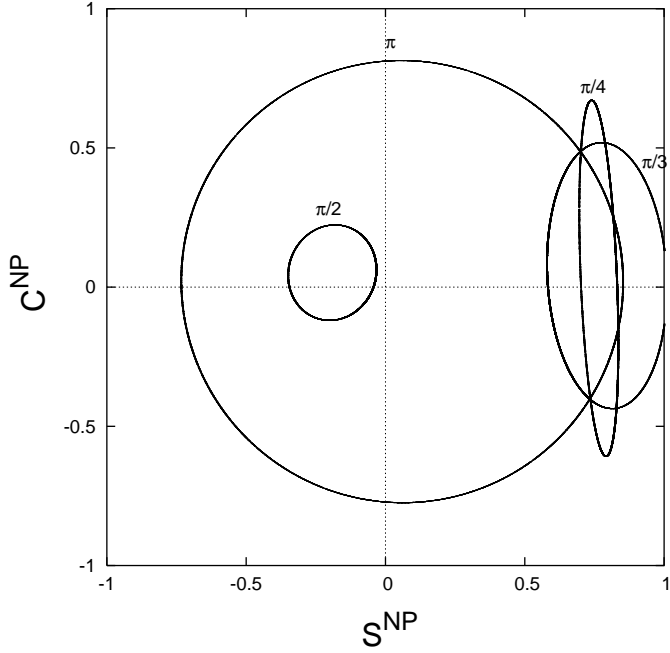}}
\caption{The  correlation plot between $S_{KK}^{NP}$ and $C_{KK}^{NP}$
for the $\B \to K^0 \bar K^0$ process in the MSSM with
$LR$ mass insertion, where we have used $r=2.6$, $\delta=11^\circ$,
$(r_{NP})_{LR}=4.66$, a set of weak phases
$\theta_{N}= \pi,~\pi/2,~\pi/3,~\pi/4$, 
and varied the strong phase
$\delta_{N}$ between 0 and $2 \pi$.}
 \end{figure}



\begin{figure}[htb] 
   \centerline{\epsfysize 3.5 truein \epsfbox{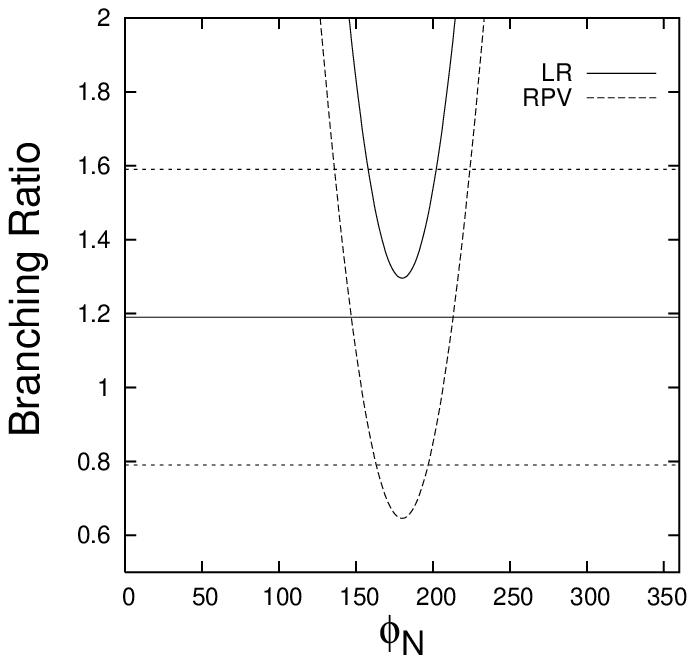}}
 \caption{
  The branching ratio of $\Bb \to K^0 \bar K^{0}$ process 
 (in units of $10^{-6}$) versus the phase
 $\phi_{N}$ (in degree). The horizontal solid line represents the 
experimental central value and the dotted lines represent the 
$1 \sigma $ range.}
  \end{figure}


\subsection{R-parity violating supersymmetric contribution}

We now analyze the decay mode in the minimal supersymmetric model with
R-parity violation (RPV). 
In the supersymmetric models there may be 
interactions which violate the baryon number $B$ and the lepton number $L$
generically. The simultaneous presence of both $L$ and $B$ number
violating operators induce rapid proton decay, which may contradict
strict experimental bound. In order to keep the proton lifetime
within the experimental limit, one needs to impose additional symmetry
beyond the SM gauge symmetry to force the unwanted baryon and lepton
number violating interactions to vanish. In most cases this has
been done by imposing an {\it ad hoc} symmetry called R-parity defined
as, $R=(-1)^{(3B+L+2S)}$, where $S$ is the intrinsic spin of the
particles. Thus R-parity can be used to distinguish the particles ($R=+1$)
from their superpartners ($R=-1$). The conservation of R-parity implies
that the supersymmetric  particles must be produced in pairs and the 
lightest supersymmetric particle (LSP) must be stable. However, 
there is no compelling reason to require the conservation 
of R-parity. Less restrictive symmetries- conservation of baryon/lepton 
number alone can be imposed to prohibit the unwanted proton decay.
Extensive studies has been done to look for the direct as well as
indirect evidence of R-parity violation from different processes
and to put constraints on various R-parity violating couplings \cite{rpv1}.

Here, we consider only the lepton number violating effects.
The most general $R$-parity and lepton number violating
super-potential is given as
\begin{equation}
W_{\not\!{L}} =\frac{1}{2} \lambda_{ijk} L_i L_j E_k^c
+\lambda_{ijk}^\prime L_i Q_j D_k^c \;,\label{eq:eqn10}
\end{equation}
where, $i, j, k$ are generation indices, $L_i$ and $Q_j$ are
$SU(2)$ doublet lepton and quark superfields and $E_k^c$, $D_k^c$
are lepton and down type quark singlet superfields. 

Thus the relevant four fermion
interaction induced by the R-parity and lepton number violating
model is
\begin{eqnarray}
{\cal H}_{\spur{R}}&=&
-\frac{1}{2m^2_{\tilde{\nu}_i}}\eta^{-8/\beta_0 }
 \biggr[ 
   \lambda'_{i31}\lambda'^{*}_{i22}(\bar{s}_{\alpha} \gamma_{\mu} L
   b_{\beta}) (\bar{d}_{\beta} \gamma^{\mu} R s_{\alpha})
+\lambda'^{*}_{i13}\lambda'_{i22}(\bar{s}_{\alpha} \gamma_{\mu} R
b_{\beta}) (\bar{d}_{\beta} \gamma^{\mu} L s_{\alpha}) 
   \nonumber\\
&&~~~~~~~~~
 +\lambda'_{i32}\lambda'^{*}_{i12}(\bar{d}_{\alpha} \gamma_{\mu} L
b_{\beta}) (\bar{s}_{\beta} \gamma^{\mu} R s_{\alpha})
+\lambda'^{*}_{i23}\lambda'_{i21}(\bar{d}_{\alpha} \gamma_{\mu} R
b_{\beta}) 
(\bar{s}_{\beta} \gamma^{\mu} L s_{\alpha})\biggr], 
\end{eqnarray}
where $\eta=\frac{\alpha_{s}(m_{\tilde{f}_i})}{\alpha_{s}(m_b )}$ and
$\beta_0 =11-\frac{2}{3}n_f $. The QCD correction factor
$\eta^{-8/\beta_0 }$ arises due to running from the sfermion mass 
scale $m_{\tilde{f}_i}$ (100 GeV assumed) down to the
$m_b$ scale.

The amplitude for $\B \to K^0 \bar K^0 $ process in the RPV model
is given as
\begin{eqnarray}
 A_{\spur{R}}(B^0 \to K^0 \bar K^0)&=& 
-\frac{1}{8 m^2_{\tilde{\nu}_i}} \eta^{-8/\beta_0 }X
\biggr[ \frac{1}{N}\left( \lambda'_{i22}
{\lambda'}^{*}_{i13}-\l'_{i31}{\l'}^*_{i22} \right )\nn\\
&-& \frac{2 m_K^2}{(m_b-m_s)(m_d+m_s)}
\left (\lambda'_{i21} {\lambda'}^*_{i23} 
-\l'_{i32} {\l'}^*_{i12} \right )\biggr]\;,
\end{eqnarray}
where we have kept only the leading order factorization contributions.
We use the parameterization $\lambda'_{i22}
{\lambda'}^{*}_{i13}=-\l'_{i31}{\l'}^*_{i22} =k e^{i \theta }$
and $\lambda'_{i32} {\lambda'}^*_{i12}=- 
\l'_{i21} {\l'}^*_{i23} =k_1 e^{i \theta}$, assuming the same
weak phase for all the RPV couplings.
The limits on the couplings $|\lambda'_{i32} {\lambda'}^*_{i12}|= 
|\l'_{i21} {\l'}^*_{i23}|$ are obtained from
$\B \to \phi \pi $ decay in Ref. \cite{eilam} 
\be
k_1=|\lambda'_{i32} {\lambda'}^*_{i12}|= 
|\l'_{i21} {\l'}^*_{i23}|\leq 4.0 \times 10^{-4}\;.
\ee
In our analysis we use $k=k_1 \leq 4.0 \times 10^{-4}$
 and obtain the new physics parameter 
\be
r_{NP}\leq 3.92 \;.
\ee
The correlation plot between $C_{KK}^{NP}$ and $S_{KK}^{NP}$
for the above value of $r_{NP}$ is shown in Figure-5, for
some representative values of the weak phase and $0 \leq
\delta^{NP} \leq 2 \pi $. The values of $r$ and $\delta$ are used as
derived from QCD factorization. In this case also one can get 
observable CP violation effects. 
Plotting the branching ratio (\ref{br11}) vs. $\phi_{N}$ 
for $|A_{NP}/A_{SM}|=1.84$ we can 
see from Figure-4 that
the observed branching ratio can be easily accommodated in the RPV model.

\begin{figure}[htb] 
 \centerline{\epsfysize 3.5 truein \epsfbox{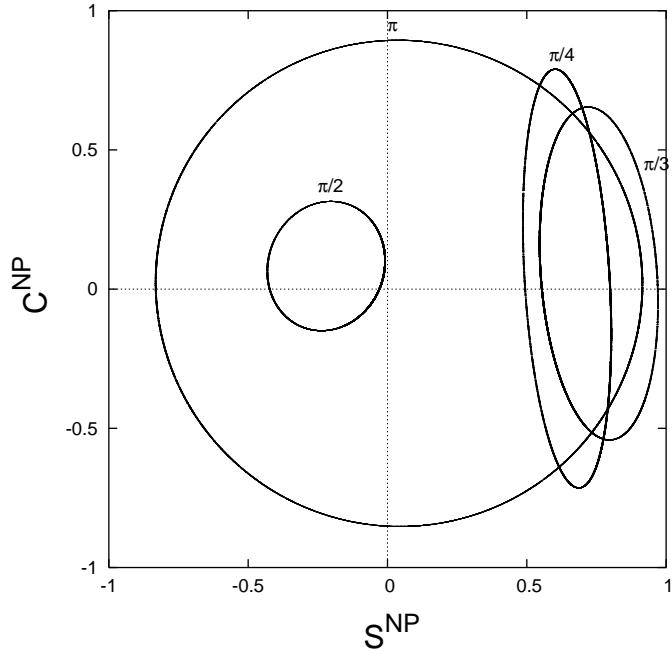}}
\caption{The  correlation plot between $S_{KK}^{NP}$ and 
$C_{KK}^{NP}$ in the RPV model for $r_{NP}=3.92$, 
$r=2.6$, $\delta=11^\circ$, $\theta_{N}=\pi,~ \pi/2, 
~\pi/3,
~\pi/4$ and
$0\leq \delta_{N} \leq 2 \pi $}
 \end{figure}


\section{Conclusion}
The recent measurement of the mixing induced CP asymmetry in $\B \to 
\phi K_S$ which has significant deviation from $\sin(2 \beta)_{\psi K_S}$ 
may provide the first indication of new physics effects present
in the $ b \to s$ penguin amplitudes. In this paper, we have investigated
the possibility of observing new physics effects in the $b \to d $ 
penguin amplitudes. We have considered the decay mode $\B \to 
K^0 \bar K^0 $ which proceeds through the quark level FCNC 
transition $b \to d \bar s s$, receiving contributions only 
from one-loop $ b \to d $ penguin diagrams. If one would assume
only the {\it top} quark exchange in the penguin loop as usually done, 
the CP asymmetry
parameters would vanish in the SM. However, contributions from 
penguins with internal {\it up} and {\it charm} quark 
exchanges are expected to yield
small non-vanishing CP asymmetries. Thus, if significant CP asymmetries 
will be found in this channel then it would be a clear indication of 
new physics effects in $b \to d $ penguin amplitudes. However, as discussed
in Ref. \cite{ciu01}, the nonfactorizable long-distance $charm$  
penguins may also give significant contributions which 
in turn yield sizable
CP asymmetries. In that case it is practically 
impossible to disentangle the
new physics effects from the nonfactorizable {\it charm} 
and {\it GIM} penguins without
any additional assumptions. However, very recently, it  has
been pointed out by Beneke {\it et al} \cite{mn04} that the nonfactorizable
$charm$ penguin contributions are of higher order in $1/m_b$ expansion.
Thus the observation of sizable CP asymmetry in this mode may be considered 
as the signal of new physics.

Using QCD factorization approach, we found the CP averaged
branching ratios in the SM for $\B \to K^0 \bar K^0 $ process
as $\sim 0.9 \cdot 10^{-6}$, which is slightly below the present
experimental value. The CP asymmetry 
parameters are found to be $S_{KK}=0.06$ and $C_{KK}=-0.16$. Allowing
the CKM parameters to vary within their $1 \sigma $ limits, we obtained
the allowed ranges as $0.02 \leq S_{KK} \leq 0.13$ and
$-0.17 \leq C_{KK} \leq -0.15$. If the observed values would deviate
significantly from the above ranges would be a clear signal of new physics.
We next analyzed the decay mode in the MSSM
with mass insertion approximation and found that the $LR$
insertion has significant effects than the $LL$ or $RR$ insertions.
In this case one can have significant  CP violating asymmetries.
Considering the R-parity violating supersymmetric model we found that
one can also obtain  significant CP violation
with the present available RPV couplings.
Therefore, the future experimental data on $\B \to K^0 \bar K^0$ 
CP violating parameters will
serve as a very good hunting ground for the existence of
new physics beyond the SM  and also support/rule out some of the 
existing new physics models. 

\begin{flushleft}
\begin{large}
{\bf Acknowledgments}
\end{large}
\end{flushleft}
R.M. would like to thank the HEP theory group at the Technion 
for the kind hospitality and
 AKG would like to thank Lady Davis Foundation for financial
support. The 
work of RM was partly  supported by Department of 
Science and Technology, Government of India,
through Grant No. SR/FTP/PS-50/2001.

\end{document}